\documentclass[11pt,a4paper]{article}

\usepackage{todonotes}
\setuptodonotes{inline} 

\usepackage{enumerate}
\usepackage{siunitx}
\usepackage{hyperref}
\usepackage{enumerate}
\usepackage{bm}
\usepackage[version=4]{mhchem}
\usepackage{amssymb,amsmath,amsfonts}
\usepackage{graphicx}

\newcommand{\dd}{\,\mathrm{d}}

\title{Vesicle Translocation into Closed Constrictions as a Function of Molecular Motor Parameters}

\author{Youngmin Park\footnote{Department of Mathematics, University of Florida, Gainesville, FL}
\and Thomas G. Fai\footnote{Department of Mathematics, Brandeis University, Waltham, MA} \footnote{Volen Center for Complex Systems, Waltham, MA}
}

\newcommand{\yp}[1]{{\color{black}{#1}}}

\usepackage[margin=1.25in]{geometry} 

\newcommand{\pa}{\partial}

\begin{document}

\maketitle

\begin{abstract}
We study the dynamics of molecular motor-driven transport into dendritic spines, which are bulbous intracellular compartments in neurons that play a key role in transmitting signals between neurons. We further develop a stochastic model of vesicle transport in [Park, Singh, and Fai, SIAM J. Appl. Math. 82.3 (2022), pp. 793--820] by showing that second-order moments may be neglected. We exploit this property to significantly simplify the model and confirm through numerical simulations that the simplification retains key behaviors of the original agent-based myosin model of vesicle transport. We use the simplified model to explore the vesicle translocation time and probability through dendritic spines as a function of molecular motor parameters, which was previously not practically possible.

\textbf{Relevance to Life Sciences}: We find that thinner dendritic spine geometry can greatly reduce the probability of vesicle translocation to the post-synaptic density. The cell may alter molecular motor parameters to compensate, but only to a point. These findings are consistent with the biological literature, where brain disorders are often associated with an excess of long, thin dendritic spines.


\textbf{Mathematical Content}: We use a moment-generating function to deduce that second-order moments in motor attachment times may be neglected, and therefore the first-order moment is a sufficient approximation. Using only the mean attachment times and neglecting the variance yields a tractable master equation from which vesicle mean first passage times may be computed directly as a function of geometry and molecular motor parameters.


\end{abstract}





\section{Introduction}


Dendritic spines (or simply ``spines'') are subcellular structures that densely populate dendrites of excitatory neurons in the mammalian brain. To give a sense of the sheer number of spines in the brain, consider that each of the billions of excitatory neurons in the mammalian neocortex \cite{kaas2011neocortex} receives tens of thousands of inputs on spines \cite{nimchinsky2002structure}; in other words, the number of spines in mammalian brains outnumbers the number of neurons by at least four orders of magnitude. In humans, assuming 15 billion neurons \cite{von2016search} where 80\% are excitatory \cite{watts2005excitatory}, there are an estimated 120 trillion spines, each of which exists at the interface between neurons (synapses). It is thus unsurprising that spines have been suggested to be a key ingredient of healthy brain function \cite{bloss2011evidence}, with defective spine formation implicated in a range of intellectual and cognitive developmental disorders including Autism spectrum disorder, schizophrenia, Alzheimer's disease \cite{penzes2011dendritic}, depression \cite{qiao2016dendritic}, and fragile-X syndrome \cite{lai2013structural}.

Dendritic spines are small subcellular features (spine volumes are $\SI{0.1}{\mu m^3}$ on average \cite{ofer2021ultrastructural}) and are generally well-defined structures that protrude from dendritic trunks with a thin neck that opens into a mushroom-like shape, where the top of the mushroom (called the postsynaptic density) makes contact with a presynaptic axon terminal \cite{gray1959electron}. 
We focus on a particular aspect of spine maintenance, for which both experimental data and theoretical investigations are particularly sparse: cargo vesicle transport into dendritic spines. The process is driven by molecular motor proteins that squeeze small cargo vesicles through sub-micron diameter constrictions to deliver proteins to the postsynaptic density at the spine head \cite{da2015positioning}. Experiments have shown that this movement is not always unidirectional -- it is possible for vesicles to become stuck (similar to a cork) or move away from the spine head \cite{park2006plasticity,wang2008myosin,da2015positioning}. The mechanisms underlying these directional changes are not well understood in the presence of a \textit{constriction} \cite{fai2017active,park2022coarse}. We emphasize that the constriction may induce significant drag effects on the vesicle \cite{fai2017active}, in contrast to studies of cargo transport in free space in which fluid drag may be negligible.

Several studies of motor-driven transport exhibit and analyze the existence of bidirectional movement, called the \textit{tug-of-war} effect. This effect is either included as an assumption \cite{muller2008tug,bressloff2009directed,newby2009directed,kunwar2011mechanical} or as an emergent behavior of mechanistic modeling \cite{julicher1995cooperative,newby2010random,guerin2011motion,guerin2011bidirectional,allard2019bidirectional}. However, many studies neglect fluid drag \cite{allard2019bidirectional}, or if fluid drag is included, it is incorporated using Stokes' drag law \cite{guerin2011bidirectional,smith2018assessing,bovyn2021diffusion}. While Stokes' drag law is accurate for motion in free fluid, the approximation becomes inaccurate for transport into confined and closed spaces such as those present in dendritic spines\footnote{To help clarify the meaning of ``confined and closed spaces,'' consider a single tennis ball and a tennis ball can under water. The diameter of the can is only slightly greater than the ball. If we then force the tennis ball into the can, the (incompressible) fluid can only escape by passing around the ball, parallel to the wall (it cannot escape transversely through the base or sides of the can because the plastic is impermeable). In this analogy, ``confinement'' refers to the similar diameter of the ball and can, while ``closed space'' refers to the can itself. The underlying fluid dynamics problem is virtually identical, but on a much smaller scale: the tennis ball is the vesicle and the can is the spine neck and head.}. In this context, the fluid dynamics problem changes significantly. 

We provide a brief summary of the altered fluid dynamics problem in the presence of a closed constriction (the full, detailed derivation may be found in \cite{fai2017active}). To aid in our description, we simplify the problem to capture essential features: assume that the vesicle is a rigid sphere and assume a cylindrical dendritic spine. These assumptions may seem overly artificial given that vesicles are highly deformable and that spine shapes exhibit significant diversity, but the fluid dynamics problem does not change significantly: for broad sets of spine morphologies, the vesicle is still forced into a closed space, the vesicle's diameter, even if variable in space, still remains similar to the spine neck diameter, and the vesicle is transported in an incompressible cytoplasmic fluid. With these simplifying assumptions in hand, Fai et al. (2017) observed a large velocity gradient in the narrow space between the spherical object and cylindrical domain as the vesicle is pushed towards the closed end and cytoplasmic fluid escapes towards the spine base. Here, lubrication theory applies \cite{acheson1990elementary} (Section 7.10), and the Navier-Stokes equation reduces to an ordinary differential equation. In particular, the degree of constriction is quantified by the confinement factor $\zeta$, which is mathematically equivalent to a (nontrivial) drag coefficient, where the drag due to confinement is directly proportional to the vesicle velocity \cite{fai2017active} (note that Fai et al. allowed deformable vesicles and arbitrary spine geometry). 

In recent works, we analyzed this ordinary differential equation model in significant detail, in particular by assuming a mean-field approximation of motor forces \cite{park2020dynamics}. The mean-field model allowed us to explore the effects of constriction geometries on the existence of multi-stable vesicle velocities. We also analyzed the stochastic case using a modified master equation coupled to a partial differential equation (PDE) describing the underlying (local) myosin motor positions \cite{park2022coarse}. Our work yielded predictions consistent with the experimental observations about the effect of spine geometry on vesicle transport: long, thin spines tend to encourage unidirectional motion and increase the likelihood of vesicle delivery, whereas wide, stubby spines tend to allow bidirectional motion, which can decrease the likelihood of vesicle delivery.

While previous work focused only on the \textit{spine geometry}'s effects on vesicle translocation and probability, we now explore the effects of \textit{molecular motor parameters} on vesicle translocation and probability. The key mathematical contribution involves the simplification of the coupled master equation and PDE system in \cite{park2022coarse} under biologically relevant conditions. This simplification enables analytical calculations for the average time to switch velocities in terms of individual molecular motor parameters, which in turn dramatically simplifies the calculation of vesicle translocation times and probabilities as a function of molecular motor parameters.

Insights from our modeling study are particularly valuable because this question is difficult to address experimentally. It is known that the cell may regulate molecular motor behavior by altering available levels of ATP \cite{maschi2021myosin}, utilizing chemical signaling mechanisms \cite{jaffe2005rho}, or mechanical signaling mechanisms \cite{howard2009mechanical}. However, to the best of our knowledge, the consequences of these changes on vesicle translocation times and probabilities are not known, particularly when it comes to dendritic spines: are there key molecular motor parameters that the neuron might be able to control? How do these parameters affect vesicle translocation to the neuron's detriment or benefit? These questions of control and maintenance motivate the present study.


\begin{figure}[ht!]
	\includegraphics[width=\textwidth]{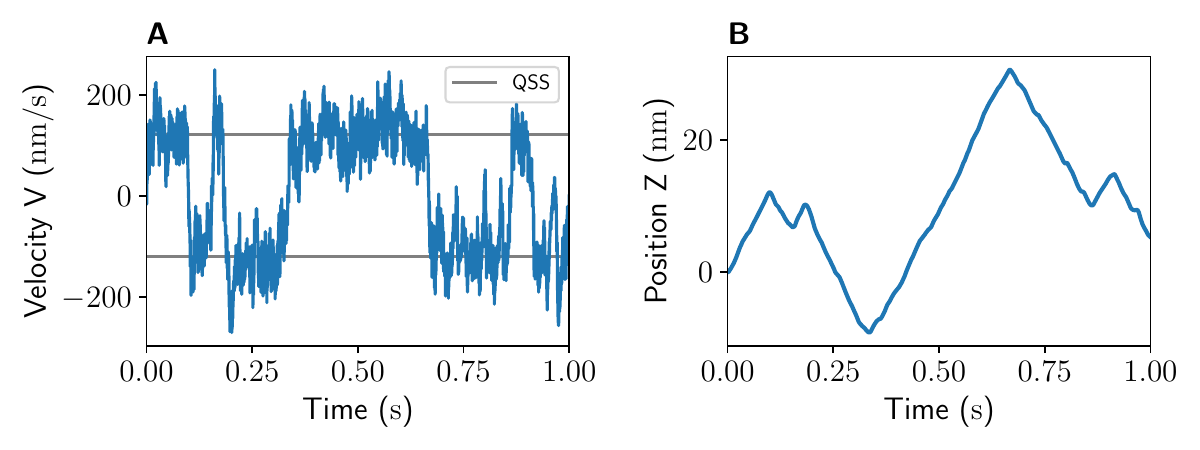}
	\caption{An example simulation of the agent-based model. A: Velocity (\si{\nm/\s}) over time (\si{\s}). Gray lines denote the quasi-steady state (QSS) velocities determined using a mean-field approximation \cite{park2020dynamics} of the agent-based model. B: Position (\si{\nm}) over time (\si{\s}). Model parameters: $N=100$, $\zeta=\SI{4e-5}{\kg/\s}$, $A=\SI{5}{\nm}$, $B=\SI{5.05}{\nm}$, $\alpha=\SI{14}{\s\tothe{-1}}$, $\beta=\SI{126}{\s\tothe{-1}}$, $p_1=\SI{4}{\pico\newton}$, $\gamma=\SI{0.322}{\nm\tothe{-1}}$. Simulation parameters: \texttt{dt=2e-6}, and total simulation time \texttt{T=1}.}\label{fig:agents}
\end{figure}

\subsection{Agent-Based Myosin Motor Model}\label{sec:agents}
We focus on using myosin motors (as opposed to kinesin or dynein) because experimental evidence suggests that actin-myosin interactions dominate vesicle transport in spines \cite{da2015positioning}. We use the agent-based model of myosin motors derived in \cite{park2022coarse}, which is based on the classic work of Huxley \cite{huxley1957muscle} and on more recent work by Lacker and Peskin \cite{lacker1986mathematical,hoppensteadt2012modeling}. We assume that the base of each myosin motor is permanently attached to the vesicle, while the head of each myosin motor attaches and detaches along the spine wall. The former part of this assumption simplifies the problem and is equivalent to assuming that there exists an excess of myosin motors to take the place of any motors that detach from the cargo domain. The latter part of the assumption follows from the observation that there exist actin filaments arranged along the axis of the cylinder (called \textit{linear actin} \cite{hotulainen2010actin}), and the timescale of vesicle translocation is much shorter than the timescale of actin filament growth or decay, i.e., actin filaments are effectively fixed in place.

The agent-based model consists of two identical myosin motor species that prefer to push the vesicle in opposite directions. This symmetric choice is another simplifying assumption that nevertheless captures the polarity of molecular motors while preserving potential bidirectional vesicle motion (symmetric motors acting on vesicles have been shown to yield bistable velocities with unstable zero velocity -- see \cite{julicher1995cooperative,fai2017active,allard2019bidirectional,park2022coarse} and Figure \ref{fig:agents}).

\subsubsection{Position Dynamics}

For convenience, let $\hat D$ and $\hat U$ denote the index set of \textit{attached} motors that prefer to push the vesicle downward (down motors) and upward (up motors), respectively. We allow the total number of available binding sites to differ between the two species and define $n_D$ ($n_U$) as the total number of down (up) motor binding sites.

\begin{figure}
    \centering
    \includegraphics[width=.75\textwidth]{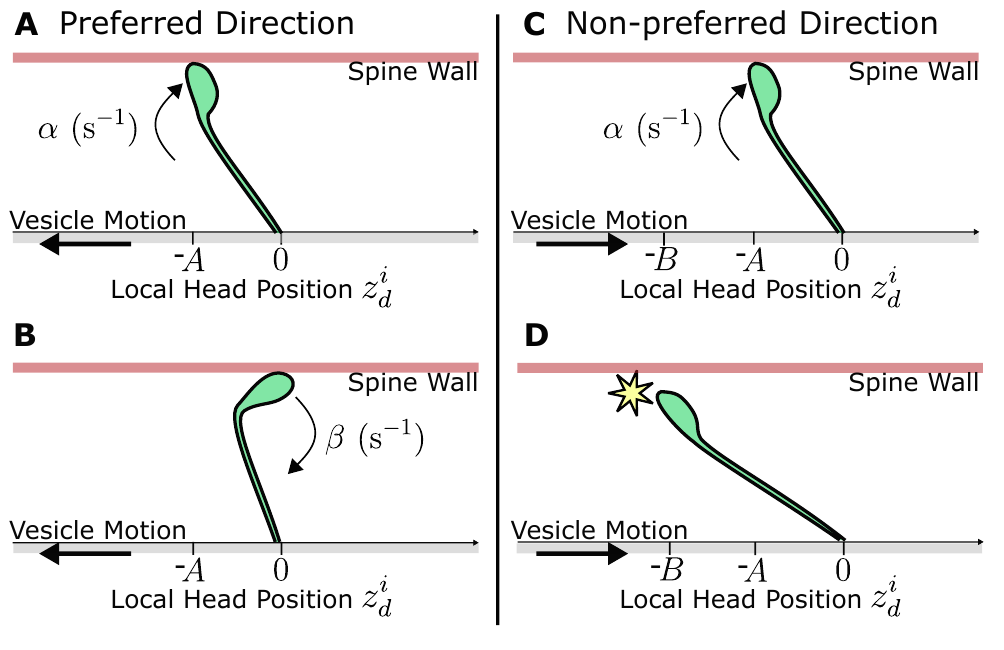}
    \caption{Myosin attachment dynamics for down motor species. A,B: When the vesicle velocity is the preferred direction $V<0$, the attachment and detachment dynamics are straightforward probabilities per unit time. C: In the non-preferred direction $V>0$, the attachment is also a simple probability per unit time. D: The detachment dynamics includes the basal rate $\beta$ combined with a forced detachment if the motor is stretched beyond its maximum extension at $z_d^i=-B$. The up motor species exhibits the same dynamics but with opposite preferred vesicle velocity and a flipped domain such that the attachment position is at $+A$ and maximum extension is at $+B$. }\label{fig:myosin}
\end{figure}

For a given vesicle velocity $V$ and time $t$, consider the index of a particular \textit{attached} motor $i \in \hat{D}$ or $i\in\hat{U}$. Local attached motor positions of the head relative to the base are denoted by $z_d^i$ and $z_u^i$, respectively, and satisfy the ODEs,
\begin{align}\label{eq:position_dynamics}
	\frac{dz_d^i}{dt} = \frac{dz_u^i}{dt} = -V,
\end{align}
where initial conditions are given by $z_d^i(0) = -A$, $z_u^i(0) = A$. The parameter $A$ is the initial attachment position of any given motor in nanometers. If a motor is \textit{unattached}, then 
\begin{equation}
    \frac{dz_d^i}{dt}=\frac{dz_u^i}{dt}=0,
\end{equation}
and $z_d^i=z_u^i=0$.

For attached motors, the domain of each local position coordinate depends on their type (up or down), due to motors having a preferred and non-preferred direction. We set the convention that the up motor species prefers positive vesicle velocity, $V>0$, while the down motor species prefers negative vesicle velocity, $V<0$. This convention results in the following domains for local motor positions conditioned on the type of motor:
\begin{equation*}
	z_d^i \in [-B\infty),
	\quad
	z_u^i \in (-\infty,B].
\end{equation*}
The parameter $B$ is the maximum amount the motor can stretch in its non-preferred direction before detaching, in nanometers. So, for example, if the position of the $i$th down motor reaches $z_d^i \leq -B$ for $V > 0$, its non-preferred direction, then this down motor will detach. Similarly, if the position of the $i$th up motor reaches $z_u^i \geq B$ for $V < 0$, its non-preferred direction, then this up motor detaches. Figure \ref{fig:myosin} summarizes the position dynamics for the down motor species.

\subsubsection{Attachment and Detachment Dynamics}
Individual down and up motors attach and detach at rates and conditions shown in Table \ref{tab:attach}. The parameters $\alpha$ and $\beta$ are the constant rates of basal attachment and detachment, respectively. When a given motor species experiences a velocity $V$ in its non-preferred direction, the rate of detachment depends on the position of the motor.
\begin{table}[ht!]
\caption{Attachment dynamics of down and up motors.}\label{tab:attach}
\begin{tabular*}{\textwidth}{c @{\extracolsep{\fill}} c}
	\hline
	Down motors & Up motors\\
        \cline{1-1} \cline{2-2}
	$\text{Unattached}\ce{ <=>[{\alpha}][{\beta\text{ and }z_d^i> -B}] }\text{Attached}$ & $\text{Unattached}\ce{ <=>[{\alpha}][{\beta \text{ and }z_u^i < B}] }\text{Attached}$
\end{tabular*}
\end{table}
A schematic of these dynamics can be seen in Figure \ref{fig:myosin} for the down motor species.

\subsubsection{Vesicle Velocity Dynamics} The vesicle velocity satisfies the force-balance equation,
\begin{equation}\label{eq:master_force}
	F_d(\{z_d^i\}) + F_u(\{z_u^i\}) - \zeta V = 0,
\end{equation}
where $F_d$ and $F_u$ are the down and up motor forces exerted by attached motors. The last term, $- \zeta V$, represents the nontrivial frictional force due to \textit{confinement} (derived in \cite{fai2017active} using lubrication theory), which is mathematically equivalent to a fluid drag effect proportional to the vesicle velocity. The motor forces are given by the sum over individual motors:
\begin{equation*}
	F_d(\{z_d^i\}) = \sum_{i\in\hat{D}}f\left(z_d^i\right), \quad F_u(\{z_u^i\}) = \sum_{i\in\hat{U}}f\left(z_u^i\right).
\end{equation*}
We neglect inertial effects due to the microscopic scale of the problem, at which viscous forces dominate.

The force-extension function  $f$ is chosen to be a linear spring,
\begin{equation*}
	f(z)=k z,
\end{equation*}
with spring constant $k$. Other choices for $f$ are possible, such as $f(z)=p_1[\exp(\gamma z)-1]$ \cite{hoppensteadt2012modeling}, but we choose a linear function for simplicity and because the choice of the force-extension function $f$ does not meaningfully change the qualitative dynamics of our problem.

In \cite{park2022coarse} it was shown that the vesicle velocity \eqref{eq:master_force} exhibits classic tug-of-war dynamics, and the effects of spine geometry on the resulting one-sided mean first passage time of vesicle translocation were explored numerically. In the present study, we turn to studying the effects of molecular motor parameters on vesicle translocation by deriving a simpler and more tractable master equation. To this end, the paper is organized as follows. In Section \ref{sec:derivation}, we derive the simplified master equation by showing, through calculations, that second-order moments of the attached motor position density may be neglected. We then compare key properties of the simplified master equation with the agent-based model in Section \ref{sec:comparisons}, in particular the steady-state distribution of quasi-stable velocities and the mean time to switch between them. Finally, we show how vesicle translocation is affected by changes in molecular motor parameters in Section \ref{sec:mfpt} and summarize biological implications of the model. We conclude with a discussion in Section \ref{sec:discussion}.

\section{Derivation of the Simplified Master Equation}\label{sec:derivation}


Let $D$ and $U$ denote the total number of attached down and up motors, respectively. To write down a master equation, we quantify the growth and decay rates of these quantities. The growth rate is straightforward to derive as $\hat \alpha_X=(n_X-X)\alpha$ for $X=D,U$, which is simply the basal attachment rate $\alpha$ scaled by the total number of unattached motors. Likewise, the decay rate in the preferred direction is given by $\delta_X = \beta X$ if $V<0$ for $X=D$ and $V>0$ for $X=U$, which is simply the basal decay rate $\beta$ scaled by the total number of attached motors.

Indeed, many models of molecular motors use constant basal growth and decay terms \cite{allard2019bidirectional}. However, our myosin model includes a position variable, which alters the detachment rate of individual motors (and therefore the population decay rate) in the non-preferred direction of each species. Thus, the vesicle velocity given by \eqref{eq:master_force} is intimately linked to the detachment rates. To make this link more explicit, let us consider a particular time $t$ and some attached motors $i\in\hat{D}$. Then each motor position satisfies the position dynamics \eqref{eq:position_dynamics},
\begin{equation}\label{eq:position}
	z_D^i(t) =  -V(t-t_D^i)-A, \quad z_U^i(t) =  -V(t-t_U^i)+A,
\end{equation}
where $t_D^{i}$ ($t_U^i$) is the attachment time of down (up) motor $i$. Then the instantaneous force exerted by each motor is given by,
\begin{equation*}
	f(z_D^i) =  -k [V(t-t_D^{i})+A], \quad f(z_U^i) = - k [V(t-t_U^{i})-A],
\end{equation*}
and the instantaneous force-balance equation becomes
\begin{align*}
	0 &=  -k\sum_{i \in \hat D} [V(t-t_D^i) + A] - k\sum_{i \in \hat U} [V(t-t_U^{i})-A] - \zeta V\\
	&= -V k\sum_{i \in \hat D} (t-t_{D}^i) - V k\sum_{i \in \hat U} (t-t_U^{i}) + kA(U-D) - \zeta V.
\end{align*}
Solving for $V$ yields
\begin{equation}\label{eq:velocity_agents}
	V = \frac{A(U-D)}{\sum_{i \in \hat D} (t-t_D^{i}) + \sum_{i \in \hat U} (t-t_U^{i}) + \zeta/k },
\end{equation}
with the consequences that we have transformed the vesicle velocity from depending on motor positions \eqref{eq:master_force} to attachment times.

However, note that Equation \eqref{eq:velocity_agents} does not provide a meaningful simplification of the agent-based model because the times at which the motors detach depend on $V$ (that is, $(t-t_D^{i})$ and $(t-t_U^{i})$ both depend on $V$ as their distributions depend on whether the velocity is in the motors' preferred direction). Thus, \eqref{eq:velocity_agents} is an implicit equation in the vesicle velocity $V$ that is quite difficult to solve explicitly. To address this, previously, the force-balance equation \eqref{eq:master_force} (equivalent to \eqref{eq:velocity_agents}) was solved numerically at each time step \cite{fai2017active}, or the local myosin motor positions were estimated using a mean-field PDE \cite{park2022coarse}.

In the present study, we take a different approach to simplify \eqref{eq:velocity_agents}: we simply replace the summation terms in the denominator with the average attachment times for each species. This substitution may appear overly simplistic, but we justify this choice by showing that second-order moments may be neglected.

\subsection{Neglecting Second-order Moments of Motor Attachment Times}\label{a:moments}

\begin{figure}[ht!]
\centering
    \includegraphics[width=.9\textwidth]{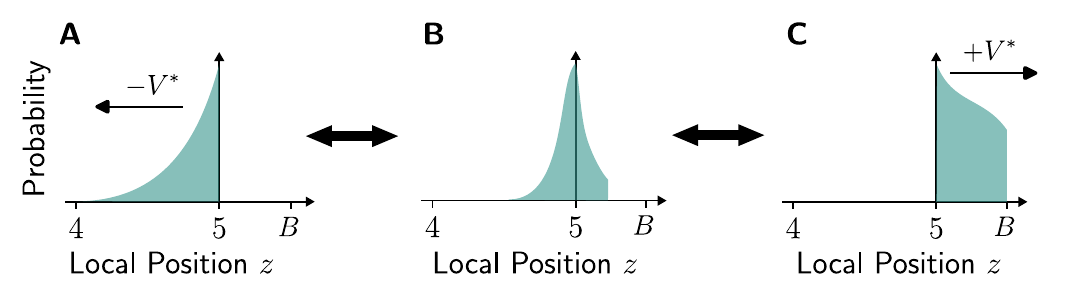}
    \caption{Schematic of densities of attached down motor (local) positions. A: When the vesicle velocity is in the preferred direction, $-V^*$, motor positions satisfy an exponential distribution ($\phi_{-V^*}^D$, Case 1). B: When the vesicle switches from $-V^*$ to $+V^*$, the density exhibits a brief time-dependent transient state ($\tilde{\phi}_{V^*}^D$, Case 2). C: The motor positions eventually reach a steady-state distribution for the new velocity $+V^*$ (${\phi}_{V^*}^D$). The velocity is now in the non-preferred direction of the down motors, so the density features an abrupt cutoff where motors are forced detach at $z = B > A = \SI{5}{\nm}$. After some time passes, the vesicle velocity switches back from $+V^*$ to $-V^*$, and the motor distributions again experience a brief transient state (Panel B, $\tilde{\phi}_{-V^*}^D$) before settling to the original steady-state density (Panel A, $\phi_{-V^*}^D$). The moments of the density in Panel A are known (they are given by the exponential distribution), whereas the moments in panels B and C require some calculations as described in  \ref{sec:transient}.}\label{fig:distribution}
\end{figure}

We make an important observation about the vesicle velocity $V$ and the  motor position dynamics that will aid in our calculations. In the agent-based model, we observe that the vesicle velocity $V$ tends to dwell about one of two quasi-stable velocities, $-V^*$ and $V^*$ from some as-of-yet unknown $V^*>0$. For concreteness, suppose the vesicle velocity $V$ is near $-V^*$ for a time such that the distributions of attached up and down motors reach a steady-state distribution $\phi_{-V^*}^U(z)$ and $\phi_{-V^*}^D(z)$, respectively (we refer to this instance as Case 1). Then, when the velocity switches from $-V^*$ to $V^*$, the distribution of attachment positions will exhibit some time-dependent transient behavior dependent on time, $\tilde\phi_{V^*}^U(z,t)$ and $\tilde\phi_{V^*}^D(z,t)$ before settling on the new respective steady-state distributions $\phi_{V^*}^U(z)$ and $\phi_{V^*}^D(z)$ (we refer to this transient state as Case 2). This observation is schematically described in Figure \ref{fig:distribution} for down motors.

We next establish that the second-order moments of the \textit{attachment times} for each species are bounded and small. Because the up and down motors are identical except in their respective preferred directions (i.e., they only differ by a linear transformation of the domain), we show our calculations for only the down motor species without loss of generality.

\subsubsection{Case 1: Steady-state}


When $V$ is near the quasi-stable velocity $-V^*$ and the attached down motors have reached an equilibrium distribution, the attachment times are exponentially distributed, so their mean and variance are simply $1/\beta$ and $1/\beta^2$, respectively \cite{hoppensteadt2012modeling,fai2017active,park2020dynamics}. However, if $V$ is near the other quasi-stable velocity  $V^*$, the vesicle is moving in the non-preferred direction of the down motors, the distribution of attachment times is less straightforward. We can nevertheless calculate the desired second-order moment using the moment generating function of the attachment time $\hat t$ for down motors conditioned on initial position $z_0=-A$:
\begin{equation}\label{eq:mgf_ss}
	\begin{split}
		\mathbb{E}[e^{\hat t q}|z_0=-A]& = \mathbb{E}[e^{T_1q}|\text{$z_0=-A$ and $z$ reaches $z=-B$}] p\\
        &\quad+ \mathbb{E}[e^{T_2 q}|\text{$z_0=-A$ and $z$ never reaches $z=-B$}](1-p),
	\end{split}
\end{equation}
where $T_1=(B-A)/V$ is the deterministic time it takes a motor position to reach $z=-B$ given the starting position $z_0=-A$, $T_2$ is an exponentially distributed random variable with rate $\beta$, and $p$ is the probability that a motor reaches $z=-B$. Note that we use the dummy variable $q$, instead of the standard $t$, to avoid potential confusion with our use of $t$ as the time variable.

If we let $f$ be the probability density function (PDF) of the exponential distribution with rate $\beta$, the probability $p$ is found by computing
\begin{equation*}
	p=1-\text{Prob}(T_2\leq T_1) = 1-\int_0^{T_1} f(s)\dd s = e^{\beta(A-B)/V}.
\end{equation*}
The conditional cumulative distribution function is given by,
\begin{equation*}
	F(T_2 | T_2\leq T_1) = \frac{\int_0^{T_2} f(s) \dd s}{\int_0^{T_1} f(s) \dd s},
\end{equation*}
and the conditional PDF is then,
\begin{equation}\label{eq:p_ss}
	f(T_2 | T_2\leq T_1) = \frac{e^{-T_2\beta}}{1-e^{-\beta(B-A)/V}}.
\end{equation}
The moment generating function $\mathbb{E}[e^{T_2 q}|\text{$z_0 = -A$ and $z$ never reaches $-B$}]$ may be computed explicitly as
\begin{align*}
	\mathbb{E}[e^{T_2 q}|\text{$z_0 = -A$ and} &\text{ $z$ never reaches $-B$}] \\
    &= \int_0^{T_1} \frac{\beta e^{-\beta s} e^{q s}}{1-e^{-\beta(B-A)/V}}\dd s = \frac{\beta  \left(e^{\frac{(B-A) (q-\beta )}{V}}-1\right)}{(\beta -q) \left(e^{\frac{\beta  (A-B)}{V}}-1\right)}.
\end{align*}
Thus, \eqref{eq:mgf_ss} may be simplified and written explicitly as the moment-generating function,
\begin{align*}
	\mathbb{E}[e^{\hat t q}|z_0=-A] = \frac{q e^{-(A-B) (q-\beta )/V}-\beta }{q-\beta }.
\end{align*}
It follows that the mean and variance of the attachment time are,
\begin{align}
	\mathbb{E}[\hat t]&= \frac{\dd}{\dd q}E[e^{\hat{t}q}|z_0=-A]|_{q=0} =
	\dfrac{1}{\beta}-\dfrac{e^{\beta(A-B)/V}}{\beta},\label{eq:moment1_steady}\\
	\mathbb{V}[\hat t]&= \frac{\dd^2}{\dd q^2}E[e^{\hat{t}q}|z_0=-A]|_{q=0} =
	\dfrac{1}{\beta^2}+\dfrac{2 \beta  (A-B) e^{\beta  (A-B)/V}-V \left(e^{2 \beta  (A-B)/V}\right)}{\beta ^2 V}\label{eq:moment2_steady}.
\end{align}

If $V<0$, then the mean is always exactly $1/\beta$ and the variance is order $O(1/\beta^2)$. If $V >0$, then the mean tends to be near $1/\beta$ and the variance is order $O(1/\beta^2)$ provided that the term $\beta (A-B)/V$ is large and negative. This negativity condition always holds because $B>A$, $\beta >0$, and $V>0$. In practice, the parameters are such that the term $\beta (A-B)/V$ is sufficiently large, thus proving boundedness of the second order. Note that the mean attachment time \eqref{eq:moment1_steady} is precisely the reciprocal of the detachment rate $\beta/[1-\exp(\beta(A-B)/V)]$ originally derived in \cite{park2022coarse}, as expected.

\subsubsection{Case 2: Transient distributions}\label{sec:transient}
As discussed in \cite{park2022coarse}, the evolution of the density of motor positions unconditioned on attachment, $\tilde \phi(z,t)$, is given by expressing the conservation of motor number through the partial differential equation,
\begin{equation}\label{eq:advection}
	\frac{\pa}{\pa t}\tilde\phi\left(z,t\right) - V\frac{\pa}{\pa z}\tilde\phi\left(z,t\right) = -\beta \tilde\phi(z,t) + \alpha(1-\theta)\delta(z+A),
\end{equation}
where
\begin{equation}\label{eq:theta0}
	\theta=\int_{-B}^\infty \tilde\phi(z,t) dz
\end{equation}
is the proportion of attached motors. To obtain $\phi(z,t)$, the density of motor positions conditioned on attachment, we simply take $\phi=\tilde\phi/\theta$. Note that if we are considering a switch in velocity from $V^*$ to $-V^*$, the mean and variance of the attachment time distribution are simply $1/\beta$ and $1/\beta^2$, respectively, because the attachment time is exponentially distributed. Therefore, we focus on the case where the velocity switches from $-V^*$ to $V^*$. In this case, we assume $\mathbb{E}[V]=V^*$, with an initial condition given by the steady-state distribution for $V=-V^*$.

First, note that for a given time $t$ and initial position $z_0$, the expected time for a motor to detach is determined by whether it first reaches the maximum extension $-B$, or detaches by the basal rate $\beta$. So, we arrive at the moment-generating function similar to \eqref{eq:mgf_ss}:
\begin{equation*}
	\mathbb{E}[e^{\hat t q}|z_0] = \mathbb{E}[e^{T_1q}|\text{$z$ reaches $-B$}] p + \mathbb{E}[e^{T_2 q}|\text{$z$ never reaches $-B$}](1-p),
\end{equation*}
but now with $z_0 \in [-A,\infty)$ (with density given by the steady-state distribution at $V=-V^*$), and where $T_1 = (B+z_0)/V$, $p=1-f(T_2\leq T_1) = e^{\beta(B+z_0)/V}$, and
\begin{equation*}
	f\left(T_2 | T_2\leq T_1\right) = \frac{e^{-T_2\beta}}{1-e^{-\beta(B+z_0)/V}},
\end{equation*}
which is found using an identical argument used to calculate \eqref{eq:p_ss}. Thus, the conditional moment generating function is
\begin{equation}\label{eq:mgf}
	\mathbb{E}[e^{\hat t q}|z_0] =\frac{q e^{\left(B+z_0\right) (q-\beta )/V}-\beta }{q-\beta }.
\end{equation}
Because $\hat t$ measures the detachment times of positions in $[-A,\infty)$, the law of total probability yields,
\begin{align*}
	\mathbb{E}[e^{\hat t q}] &= \int_{-A}^\infty \mathbb{E}[e^{\hat t q}|z]\phi(z,t)dz.
\end{align*}
Although this integral cannot be computed directly, we may bound it above by using H\"{o}lder's inequality. In our case, for smooth, integrable functions $h(z)$,$\phi(z,t)$ on the domain $[-A,\infty)$,
\begin{align*}
	\left|\int_{-A}^\infty f(z) \phi(z,t) dz\right| &\leq \|f(z)\|_\infty \int_{-A}^\infty |\phi(z,t) |dz\\
	&\leq\max_{-A\leq z <\infty}|f(z)|,
\end{align*}
where the second inequality holds because the integral of $\phi$ is at most 1 on $[-B,\infty)$ and therefore at most 1 on $[-A,\infty)\subset[-B,\infty)$. Starting with the first moment, we have
\begin{equation}\label{eq:moment1_transient}
	\begin{split}
		|\mathbb{E}[\hat t] |&\leq \max_{-A\leq z_0 <\infty}\left|\frac{1}{\beta} - \frac{e^{-\beta(B+z_0)/V}}{\beta}\right|\\
		&\leq \frac{1}{\beta} + \max_{-A\leq z_0 <\infty}\left|-\frac{e^{-\beta(B+z_0)/V}}{\beta}\right|\\
		&\leq  \frac{1}{\beta} +\frac{e^{-\beta(B-A)/V}}{\beta},
	\end{split}
\end{equation}
which follows from the triangle inequality and monotonicity of the exponential term, which decreases as a function of $z_0$ on $[-A,\infty)$. Similarly, the second moment has the bound
\begin{equation}\label{eq:moment2_transient}
	\begin{split}
		|\mathbb{E}[\hat t^2]| &\leq \max_{-A\leq z_0 <\infty}\left|\frac{2}{\beta^2} - \frac{e^{-(\beta(B+z_0))/V}(V+B\beta+\beta z_0)}{V \beta^2}\right|\\
		&\leq\frac{2}{\beta^2} + \max_{-A\leq z_0 <\infty}\left|\frac{e^{-(\beta(B+z_0))/V}(V+B\beta+\beta z_0)}{V \beta^2}\right|\\
		&\leq \frac{2}{\beta^2} + \frac{e^{-(\beta(B-A))/V}(V+\beta(B-A))}{V \beta^2}
	\end{split}
\end{equation}
Thus, in the absolute worst case, the second moment is bounded above by a term that is $O(1/\beta^2)$. Given that the parameter $\beta$ is roughly \SI{126}{s^{-1}} \cite{hoppensteadt2012modeling}, the error term $O(1/\beta^2)$ is very small, and we may safely neglect second-order moments in Case 1 and Case 2 $\Box$.

\subsection{Simplifying the Velocity Equation}

Finally, we return to the vesicle velocity derived from the agent-based model \eqref{eq:velocity_agents}. \yp{There are two cases to consider. First, for $U>D$,} we replace the somewhat complicated terms in the denominator of \eqref{eq:velocity_agents} with their mean values, yielding the equation,
\begin{equation}\label{eq:v1a}
	\yp{V_1} = \frac{A(U-D)}{D(1-e^{\beta(A-B)/\yp{V_1}})/\beta+ U/\beta + \zeta/k }.
\end{equation}
We similarly obtain the velocity equation in the case that $U<D$:
\begin{equation}\label{eq:v1b}
    \yp{V_2} = \frac{A(U-D)}{D/\beta+ U(1-e^{\yp{\beta(B-A)/V_2}})/\beta + \zeta/k }.
\end{equation}
Equations \eqref{eq:v1a} and \eqref{eq:v1b} still pose a difficulty, namely that they are transcendental in $V_i$, but they are nevertheless significantly more tractable than \eqref{eq:velocity_agents}.

While it is possible to solve \eqref{eq:v1a} and \eqref{eq:v1b} numerically, we apply one more simplification to enable analytical calculations of $V$ while maintaining relatively good accuracy compared to the original vesicle velocity. To this end, we use the $[0/1]$ Pad\'{e} approximant of the exponential:
\begin{equation*}
	\yp{e^{\beta(A-B)/V_1} \approx \frac{1}{1 - \beta(A-B)/V_1}, \quad e^{\beta(B-A)/V_2} \approx \frac{1}{1 - \beta(B-A)/V_2}.}
\end{equation*}
(See \cite{wall1999} Ch. 20 for a general treatment of $[m/n]$ Pad\'{e} approximants \yp{and see Appendix \ref{a:velocity} for a brief discussion on the accuracy on this approximation}).

We replace the exponential in \eqref{eq:v1a} and \eqref{eq:v1b} with their respective Pad\'{e} approximants to yield a pair of quadratic equations,
\begin{align*}
    a_1 V_1^2 + b_1 V_1 + c_1 = 0,\quad a_2 V_2^2 + b_2 V_2 + c_2 = 0,
\end{align*}
where $V_1$ is the velocity for $U>D$, $V_2$ is the velocity for $U<D$, and
\yp{\begin{align*}
    a_1 &= k U+\beta \zeta,\\
    b_1 &= \beta [(B-A) ((D+U)k+\beta\zeta) + Ak(D-U)],\\
    c_1 &= Ak(B-A)(D-U)\beta^2,\\
    a_2 &= k D + \beta \zeta,\\
    b_2 &= -\beta [(B-A) ((D+U)k+\beta\zeta) + Ak(U-D)],\\
    c_2 &= Ak(B-A) (U-D)\beta^2.
\end{align*}}
The solutions are given by,
\begin{equation*}
    V_1 = \frac{-b_1 \pm \sqrt{b_1^2 -4a_1c_1}}{2a_1}, \quad V_2 = \frac{-b_2 \pm \sqrt{b_2^2 -4a_2 c_2}}{2a_2}.
\end{equation*}
\yp{To select the correct root for $V_1$, we select the root corresponding to $V_1>0$ because $U>D$. Similarly, we select the root corresponding to $V_2<0$ because $U<D$. Thus,
\begin{equation}\label{eq:v_master}
    V = \begin{cases}
        V_1 = \dfrac{-b_1 + \sqrt{b_1^2 -4a_1c_1}}{2a_1} & \text{if } U>D,\\
        0 & \text{if } U = D,\\
        V_2 = \dfrac{-b_2 - \sqrt{b_2^2 -4a_2c_2}}{2a_2} & \text{if } U<D.
    \end{cases}
\end{equation}
Equation \eqref{eq:v_master} is the velocity used in the simplified master equation.

To confirm the choices of the roots of $V_1$ and $V_2$, note that $a_1 > 0$ and $c_1 < 0$ by inspection. It follows that
\begin{align*}
    &b_1^2 - 4a_1 c_1 > b_1^2 \\
    &\Rightarrow -b_1 + \sqrt{b_1^2 -4 a_1 c_1} > -b_1 + |b_1| \geq 0,
\end{align*}
from which it follows that $V_1>0$. A similar argument can be used to show that $V_2 < 0$.
}

Note that we leave the detachment rate in the non-preferred direction unsimplified,
\begin{equation}\label{eq:gamma}
	\gamma(V)=\frac{\beta}{1-\exp(-\beta(B-A)/|V|)},
\end{equation}
because simplifying \eqref{eq:gamma} in the same way as the velocity equation -- using Pad\'e approximants -- results in the loss of desired fixed points.

Now that we have obtained a simplified velocity equation that depends directly on the number of attached motors, we are able to perform analytical calculations using the simplified master equation and obtain its steady-state probability density and the mean time to switch vesicle velocities.

\subsection{Probability Density Equation}
Let $P_{D,U}(t)$ be the occupation probability of $D$ down motors and $U$ up motors attached at time $t$, for $0\leq D \leq n_D$ and $0\leq U \leq n_U$. The occupation probability satisfies the differential equation,
\begin{equation}\label{eq:dp1}
	\begin{split}
		\frac{\pa P_{D,U}}{\pa t} &= -\hat\delta_{D,U} P_{D,U}  + \hat\alpha_{D-1} P_{D-1,U} + \hat\alpha_{U-1}P_{D,U-1} \\
		&\quad + \delta_{D+1,U}^D P_{D+1,U} + \delta_{D,U+1}^U P_{D,U+1},
	\end{split}
\end{equation}
where $\hat\delta_{D,U} = \delta^D_{D,U}+\delta^U_{D,U} +\hat\alpha_D + \hat\alpha_U $ is the sum of all transition rates leaving state $(D,U)$, $\hat \alpha_X=(n_X-X)\alpha$ for $X=D,U$ is the population growth rate, and the population detachment rate is given by,
\begin{equation*}
	\delta_{D,U}^X = \begin{cases}
		\begin{cases}
			D\beta &\text{if $D\geq U$}\\
			\frac{D\beta}{1-\exp(\beta(A-B)/|V(D,U)|)} &\text{if $D<U$}
		\end{cases}& \text{if $X=D$}\\\\
		\begin{cases}
			\frac{U\beta}{1-\exp(\beta(A-B)/|V(D,U)|)} &\text{if $D>U$}\\
			U\beta &\text{if $D\leq U$}\\
		\end{cases}& \text{if $X=U$}
	\end{cases}
\end{equation*}
We use a lexicographical ordering to form a vector of all occupation probabilities,
\begin{equation*}
	\bm{P} = [P_{00}, P_{10},\ldots P_{n_D,0},P_{01},P_{11},\ldots,P_{n_D,1},\ldots,P_{n_D,n_U}]^T,
\end{equation*}
where an arbitrary index $I$ of $P$ is given by $I = D+(n_D+1)U$, and its inverse is given by $(D,U)=(I\mod(n_D+1),\lfloor I/(n_U+1)\rfloor)$. Then we may write \eqref{eq:dp1} in vector form,
\begin{equation}\label{eq:dq}
	\frac{d\bm{P}}{dt}= \bm{Q}^T \bm{P},
\end{equation}
where $\bm{Q}$ is a generator matrix consisting of the right-hand side terms in \eqref{eq:dp1}. The generator matrix $\bm{Q}$ is $(n_Dn_U)\times(n_Dn_U)$ block tri-diagonal:
\begin{equation}\label{eq:qt}
	\bm{Q}^T = \left(\begin{matrix}
		\bm{M}_0 & \bm{D}_1 & \bm{O} & \cdots\\
		\hat \alpha_0 \bm{I}_{n_D} & \bm{M}_1 & \bm{D}_2 & \bm{O} & \cdots\\
		\bm{O}& & \ddots & \\
		\cdots & \bm{O} & \hat \alpha_{n_U-2} \bm{I}_{n_D} & \bm{M}_{n_U-1} & \bm{D}_{n_U} \\
		& \cdots & \bm{O} & \hat \alpha_{n_U-1} \bm{I}_{n_D} & \bm{M}_{n_U}
	\end{matrix}\right),
\end{equation}
where $\bm{M}_j$ is tri-diagonal and $\bm{D}_j$ is diagonal:
\begin{equation}\label{eq:md}
	\bm{M}_j = \left(\begin{matrix}
		-\hat\delta_{0,j} & \delta^D_{1,j} \\
		\hat \alpha_0 & -\hat\delta_{1,j} & \delta^D_{2,j}& \\
		&\ddots &  & \\
		& \hat \alpha_{n_D-2} & -\hat\delta_{n_D-1,j} & \delta^D_{n_D,j} \\
		&  & \hat \alpha_{n_D-1} & -\hat\delta_{n_D,j}
	\end{matrix}\right),
	\bm{D}_j = \left(\begin{matrix}
		\delta^U_{0,j} & \\
		&  &  & \\
		&\ddots &  & \\
		& &  &\\
		&  & & \delta^U_{n_D,j}
	\end{matrix}\right).
\end{equation}
Note that $n_D$ determines the size of each block of $\bm{Q}$ and $n_U$ determines the number of blocks in $\bm{Q}$. Thus, even if $n_D \neq n_U$, the generator matrix $\bm{Q}$ remains square and this formulation remains applicable. The block tridiagonal generator matrix $\bm{Q}$ is called a level-dependent quasi birth-death process (LDQBD) in the literature \cite{gaver1984finite,Bright1995,bean2000quasistationary,Baumann2010,baumann2013computing,mandjes2016running,Cordeiro2019}.

\section{Comparison of the Simplified Master Equation and Agent-based Model}\label{sec:comparisons}

Now that we have derived the simplified master equation \eqref{eq:dq}, we turn to calculating the steady-state distributions and the mean time to switch velocities for \eqref{eq:dq} and compare the corresponding quantities in the agent-based model. This numerical verification serves to further confirm that we may neglect second-order moments as shown in Section \ref{a:moments}.

\subsection{Steady-State}\label{sec:steady-state}
The existence of stationary states for LDQBDs is well-established \cite{gaver1984finite,Bright1995,Baumann2010,baumann2013computing,Cordeiro2019}. The generator matrix $\bm{Q}$ corresponds to a strongly connected graph, which implies that $\bm{Q}$ is irreducible and there exists a unique steady-state $\bm{P}^*$ satisfying
\begin{equation}\label{eq:master_steady}
	\begin{split}
		\bm{Q}^T \bm{P}^* &= 0,\\
		P^*_{00} + P^*_{10} + \cdots + {P}^*_{n_D,n_U} &= 1.
	\end{split}
\end{equation}
This system may be solved rapidly using a sparse solver, because the generator matrix $\bm{Q}$ is sparse. In contrast, obtaining a steady-state distribution in the agent-based model requires long simulation times to ensure convergence \cite{park2022coarse}. 

\begin{figure}
	\centering
	\includegraphics[width=\textwidth]{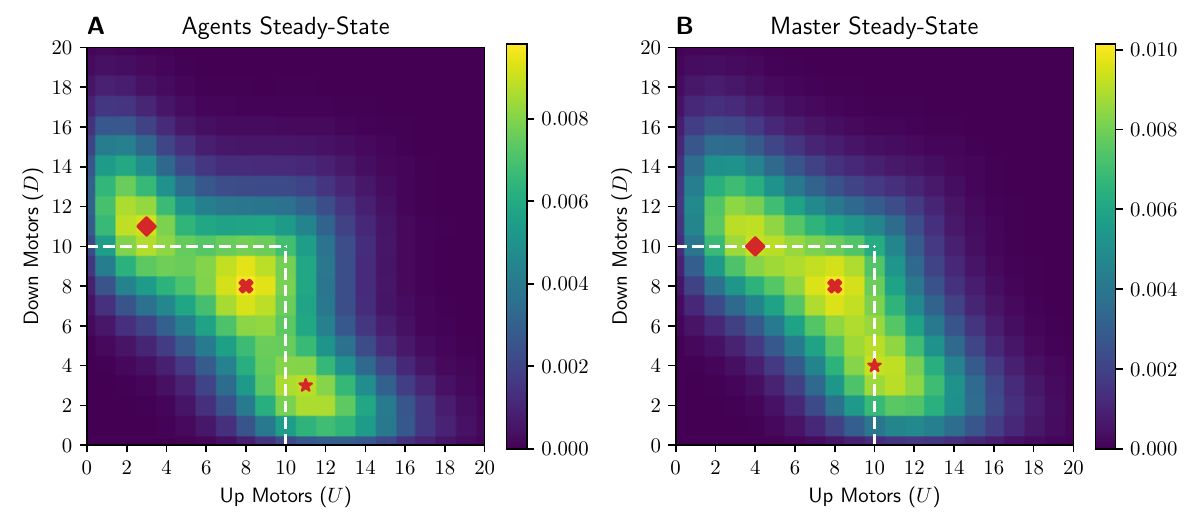}
	\caption{Comparison between the steady-state distribution in the agent-based model (A) and simplified master equation (B). Parameters: $n_D = n_U = 100$, $A=5$, $B=5.04$, $\alpha=14$, $\beta=126,\zeta=3.1$. White dashed lines denote the upper bound for the steady-state peaks in the Fokker-Planck equation. To compare the steady-state peaks as a function of $\zeta$, we track the number of attached down motors corresponding to each peak (see Figure \ref{fig:ss}). }\label{fig:ss_example}
\end{figure}

A comparison of the steady-state distribution of the agent-based model and the master equation is shown in Figure \ref{fig:ss_example} using the same representative parameters. The parameters are chosen so that there are three peaks in the distribution of attached motors for each panel/model. Each peak in the distribution corresponds to a negative vesicle velocity (red diamond), a zero vesicle velocity of the vesicle (red x) and a positive vesicle velocity (red star). These distributions are strikingly qualitatively similar, at least for the chosen set of parameters, providing initial confirmation that the choice to neglect second-order moments is valid.

To further verify the accuracy of the simplified master equation, we track the steady-state peaks as a function of the constriction factor $\zeta$. This task is computationally straightforward because we only need to solve a sparse system of equations \eqref{eq:master_steady}. This calculation yields a significant time improvement in comparison to the agent-based model, which requires a total recalculation of the distribution for each incremental change in the constriction factor $\zeta$.

We next consider the difficulties associated with the detection of peaks in the steady-state distributions. Peak-detection methods are prone to errors, especially near bifurcations: as peaks begin to merge, their shapes become poorly defined. We have no good alternative peak detection strategy within the agent-based model and must accept that peaks detected near bifurcations may be inaccurate or spurious. The same errors apply to detecting peaks in the simplified master equation. Even though the steady-state distributions computed using \eqref{eq:master_steady} are theoretically sound and have relatively well-defined peaks, these peaks become difficult to detect near bifurcations. However, we have an option to mitigate this issue in the simplified master equation.

For the purpose of tracking peaks in the steady-state distribution, we may further simplify the simplified master equation to enable the use of numerical continuation using XPPAUTO \cite{xpp}. This simplification comes at the cost of a more accurate comparison with the agent-based model, but it comes with the benefit of accurately tracking peaks even arbitrarily close to bifurcations. This simplification is only possible due to the significant tractability of the simplified master equation.

We perform the further simplification (which is used only for purposes of peak detection) by deriving the Fokker-Planck equation and identifying the points in phase space with vanishing drift. Let $x=D/n_D$, $y=U/n_U$, $\ell_x=1/n_D$, and $\ell_y=1/n_U$. Then for some sufficiently smooth function in two variables, say $g(x,y)$, the relevant Taylor expansions to second order are given by,
\begin{align*}
	g(x\pm \ell,y) &\approx g(x,y) \pm \ell_x\pa_x g(x,y) + \frac{\ell^2}{2}\pa_x^2g(x,y),\\
	g(x,y\pm \ell) &\approx g(x,y)\pm\ell_y\pa_y g(x,y) + \frac{\ell^2}{2}\pa_y^2g(x,y).
\end{align*}
Truncating the master equation up to second order yields
\begin{equation}\label{eq:fp}
	\frac{\pa P(\bm{z})}{\pa t} = -\nabla \cdot ([\bm{\alpha}(\bm{z})-\bm{\delta}(\bm{z})]P(\bm{z})) + A \cdot ([\bm{\alpha}(\bm{z})+\bm{\delta}(\bm{z})]P(\bm{z})),
\end{equation}
where $\bm{z} = (x,y)$, $A:=\frac{1}{2}(\ell_x^2 \pa_x^2,\ell_y^2\pa_y^2)$ is a differential operator (if $\ell_x = \ell_y$ then $A$ is the Laplace operator), and vectors $\bm{\alpha}$ and $\bm{\delta}$ consist of the attachment and detachment rates:
\begin{align*}
	\bm{\alpha}(\bm{z})&=(\hat\alpha^D(x),\hat\alpha^U(y))^T,\\
	\bm{\delta}(\bm{z})&=(\delta^D(\bm{z}),\delta^U(\bm{z}))^T.
\end{align*}
Suppose, without loss of generality, that there are more attached down motors than up motors, so $x>y$. The relevant term of \eqref{eq:fp} is the drift term $\bm{\alpha}(\bm{z})-\bm{\delta}(\bm{z})$, which is the deterministic part of the corresponding Langevin equation. At equilibrium, $\bm{\alpha}(\bm{z}^*)-\bm{\delta}(\bm{z}^*) = 0$, which yields the system,
\begin{align}
	\alpha(1-x^*) &=\beta x^*,\label{eq:x_eq}\\
	\alpha(1-y^*) &=\frac{\beta y^*}{1-\exp(\beta(A-B)/V(x^*,y^*))}\label{eq:y_eq},
\end{align}
where $V(x^*,y^*)$ is the equilibrium vesicle velocity. The first equation is trivial to solve,
\begin{equation}\label{eq:x_expression}
	x^*=\alpha/(\alpha+\beta),
\end{equation}
which represents the proportion of attached down motors in their preferred direction ($V<0$) at equilibrium. The second equation may be written
\begin{equation}\label{eq:y_expression}
	\begin{split}
		y^* &= \frac{\alpha(1-\exp(\beta(A-B)/V(x^*,y^*)))}{\beta+\alpha(1-\exp(\beta(A-B)/V(x^*,y^*))}\\
		&=\frac{\alpha(1-\exp(\beta(A-B)/V))+\beta - \beta}{\beta+\alpha(1-\exp(\beta(A-B)/V(x^*,y^*)))}\\
		&=1-\frac{\beta}{\beta+\alpha(1-\exp(\beta(A-B)/V(x^*,y^*)))},
	\end{split}
\end{equation}
which represents the proportion of attached down motors in their non-preferred direction. By solving \eqref{eq:y_expression} we may approximate the peaks of the steady-state distributions.

To solve \eqref{eq:y_expression} numerically, recall that we had simplified the velocity equation to explicitly depend on the number of attached up and down motors derived using Equations \eqref{eq:v1a} and \eqref{eq:v1b}. By replacing $V(x^*,y^*)$ in \eqref{eq:y_expression} with $V(x^*n_D,y^*n_U;\zeta)$, we obtain the steady-state equation,
\begin{equation*}
		y^* = 1-\beta[\beta+\alpha(1-\exp(\beta(A-B)/V(x^*n_D,y^*n_U;\zeta)))]^{-1}.
\end{equation*}
We now use numerical continuation to solve this equation as a function of the constriction factor $\zeta$.

\begin{figure}[ht!]
	\includegraphics[width=\textwidth]{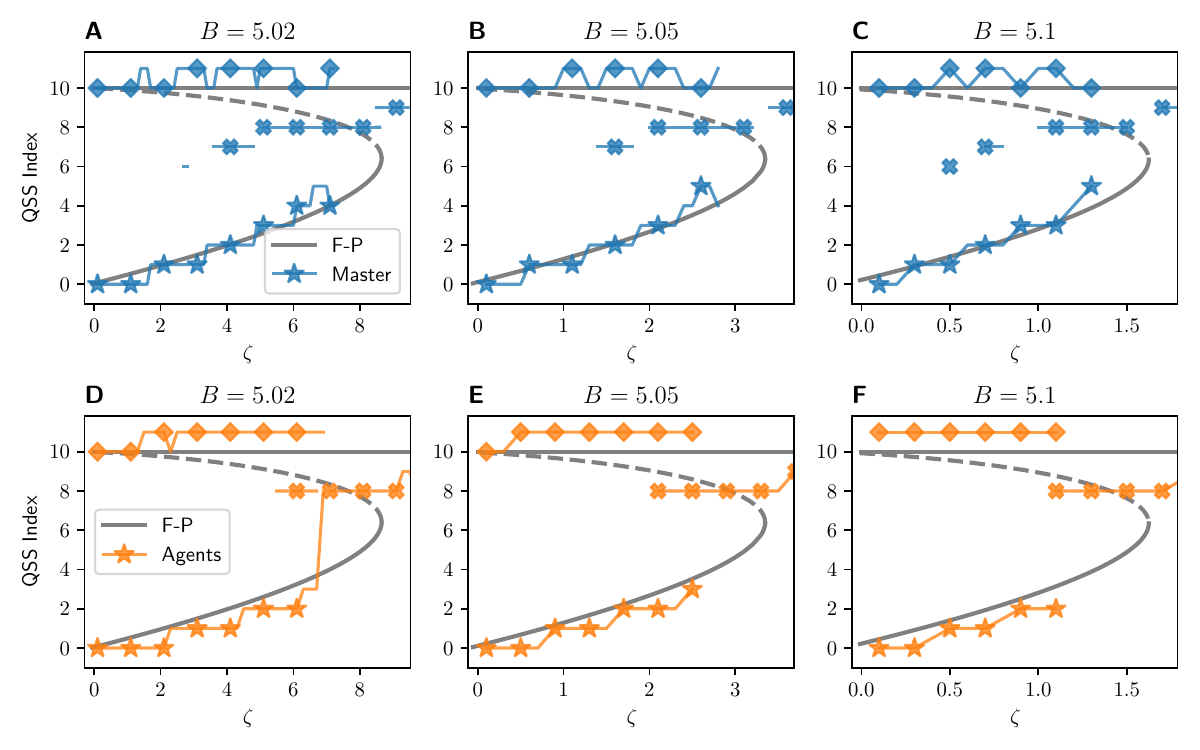}
	\caption{Steady-states in the simplified master equation (top row, blue) and agent-based model (bottom row, orange) compared to the Fokker-Planck (F-P) equation (gray). The down motors y-axis label corresponds to the number of attached down motors while the system is at a quasi steady-state equilibrium. The qualitative behavior of both models are nearly identical as a function of $\zeta$. For small values of $\zeta$, there is an upper branch (marked with diamonds) and lower branch (marked with stars), corresponding to the two peaks in the steady-state distribution. For greater values of $\zeta$, a middle branch appears (marked with ``x''s). For sufficiently large $\zeta$, only the middle branch persists.}\label{fig:ss}
\end{figure}

To simplify the presentation, we only plot the number of down motors in each peak of the steady-state distribution for each model (Figure \ref{fig:ss}). The top row of Figure \ref{fig:ss} (panels A--C) shows the estimated peaks in the simplified master equation (blue) overlaid on the peaks obtained using the corresponding Fokker-Planck equation (gray). The blue symbols (diamond, x, and star) correspond to the top left, middle, and bottom right peaks shown in Figure \ref{fig:ss_example}. In Figure \ref{fig:ss}A, the Fokker-Plank (F-P) steady-state peaks (gray) closely match the location of the steady-state peaks of the simplified master equation (blue) as a function of constriction factor $\zeta$, especially for looser constrictions. However, as the constriction factor $\zeta$ increases, we lose the ability to accurately estimate peaks in the simplified master equation, whereas the peaks computed using the F-P equation are accurately tracked through the bifurcation where the peaks merge for $\zeta$ sufficiently large. Beyond this bifurcation point, only the zero velocity exists in both the F-P formulation and the simplified master equation.

Figure \ref{fig:ss}A, B, C correspond to different values of the detachment position $B \in\{\SI{5.02}{nm}, \SI{5.05}{nm}, \SI{5.1}{nm}\}$, respectively. Note that the $x$-axis scale differs significantly in panels A--C despite sub-nanometer differences in the initial attachment position; these different scales indicate that the F-P formulation correctly captures the qualitative bifurcation points. We explore additional consequences of altering the detachment position in Section \ref{sec:mfpt}, in which we look at how molecular motor parameters affect the vesicle translocation time and probability.


The bottom row of Figure \ref{fig:ss} (panels D--E) shows the same information as the top row (panels A--C), but for the agent-based model (orange). In conclusion, we find that the F-P equation, the simplified master equation, and the agent-based model qualitatively agree on the existence and disappearance of steady-state peaks for various parameter values. With this similarity established, we now turn our attention to verifying the qualitative agreement in the mean time to switch velocity between the simplified master equation and the agent-based model in Section \ref{sec:v_mfpt}. This quantity is important because it affects the probability and mean time to vesicle translocation, which we will analyze in Section \ref{sec:mfpt}.


\subsection{Mean Time to Switch Velocity}\label{sec:v_mfpt}

We avoid using the continuum approximation (the Fokker-Planck equation \eqref{eq:fp}) to compute the mean time to switch velocity because such mean first passage time calculations of birth-death processes are known to disagree between the continuum limit and the deterministic model \cite{doering2005extinction,park2022coarse}. However, the mean time to switch velocity is straightforward to calculate using a master equation \cite{norris1997markov,allard2019bidirectional,grimmett2020probability}, including our simplified master equation. In fact, there is only one calculation required, namely the time to switch from a given state $I$ (representing some number of $(D_I,U_I)$ attached motors) to another given state $J$ (representing some number of $(D_J,U_J)$ attached motors).

To this end, let $\bm{Q}_J$ denote a generator matrix with row and column $J$ removed, which is equivalent to $\bm{Q}_J$ corresponding to the generator matrix of a graph where $J$ is an absorbing state. Survival probabilities are given by
\begin{equation*}
	\bm{P}_J(t)(1,\ldots,1)^T = \exp(t\bm{Q}_J)(1,\ldots,1)^T,
\end{equation*}
and the distribution of hitting times is the negative derivative of the above. If we define the vector of expected hitting times from state $I\neq J$ to $J$ as
\begin{equation*}
	\bm{\tau}_J = (\tau_{1,J},\ldots,\tau_{J-1,J},\tau_{J+1,J},\ldots,\tau_{n_Dn_U,J}),
\end{equation*}
then
\begin{equation}\label{eq:tauj}
	\begin{split}
	\bm{\tau}_J &= \int_0^\infty -t \frac{d}{dt} \bm{P}_J(t)(1,\ldots,1)^T dt\\
	&= \int_0^\infty \exp(t\bm{Q}_J)dt (1,\ldots,1)^T\\
	&= -\bm{Q}_J^{-1} (1,\ldots,1)^T.
	\end{split}
\end{equation}
We now have an explicit equation relating the average time to switch velocity to all model parameters, which are contained in $\bm{Q}_J$. Equation \eqref{eq:tauj} therefore yields an analytical formula for computing the average time to switch velocity. The term $\tau_{I,J}$ is the average time to switch from state $I$ to state $J$ (in the lexicographical ordering), with $I$ and $J$ chosen such that they each correspond to the index of peaks in the steady-state distribution. For simplicity, we will drop subscripts and denote
\begin{equation*}
    \tau \equiv \tau_{I,J},
\end{equation*}
with the understanding that $\tau$, in addition to depending on the initial and final states $I$ and $J$, directly depends on the microscopic motor parameters $\alpha$ (basal attachment rate), $\beta$ (basal detachment rate), $A$ (local attachment position), $B$ (local detachment position), $k$ (spring constant), $n_D$ (total number of available down motors), $n_U$ (total number of available up motors), and $\zeta$ (confinement factor).

\begin{figure}[ht!]
	\centering
	\includegraphics[width=\textwidth]{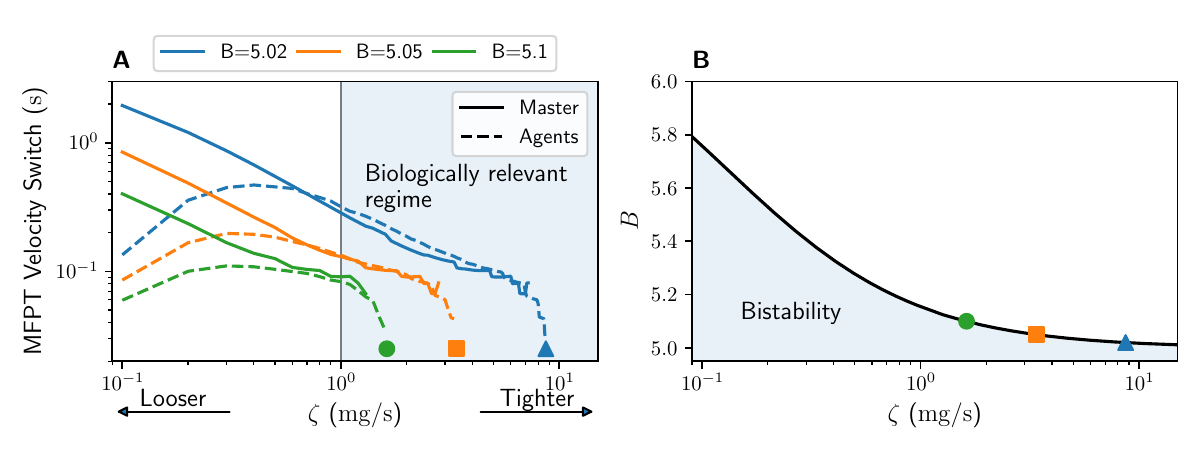}
	\caption{Average time to switch velocity. A: Simplified master equation (solid) and agent-based model (dashed) for $B=\SI{5.02}{nm}$ (blue), $B=\SI{5.05}{nm}$ (orange), and $B=\SI{5.1}{nm}$ (green). The y-axis labeled ``MFPT Velocity Switch (\si{s})'' is $\tau$ for the master equation. The respective blue triangle, orange square, and green circle denote the constriction factor ($\zeta$) values where the bimodal steady-state peaks merge into a single peak as predicted by the Fokker-Plank equation. After the peaks merge, it is no longer possible to compute the mean time to switch velocity, because there is only one stable velocity. The blue shaded region corresponds to the biologically relevant regime estimated in \cite{park2020dynamics}. B: Two-parameter bifurcation diagram of the Fokker-Planck equation. The black curve corresponds to parameter values where the pair of steady-state peaks merge into one. The blue triangle, orange square, and green circle correspond directly to the same shapes in Panel A. The blue shaded region corresponds to parameter values where two peaks exist in the steady-state distribution of the Fokker-Planck equation. Parameters: $n_D = n_U = 100$, $A=\SI{5}{nm}$, $\alpha=\SI{14}{\s^{-1}}$, $\beta = \SI{126}{\s^{-1}}$.}\label{fig:v_switch}
\end{figure}

Comparisons of the average time to switch velocity in the simplified master equation and the agent-based model are shown in Figure \ref{fig:v_switch}. In panel A, the mean first passage time to switch velocity in the master equation (black) and agent-based model (dashed) agree as a function of the constriction factor $\zeta$, at least in the biologically relevant regime (blue region). Panel B shows a two-parameter bifurcation diagram of the detachment position $B$ and the confinement factor $\zeta$. Here, the blue region corresponds to bistability, where there exist two quasi-stable velocities. In summary, Figure \ref{fig:v_switch} shows that the mean time to switch vesicle velocity can be highly sensitive to microscopic motor parameters, in this case the detachment position $B$.

\begin{figure}[ht!]
    \centering
    \includegraphics[width=\textwidth]{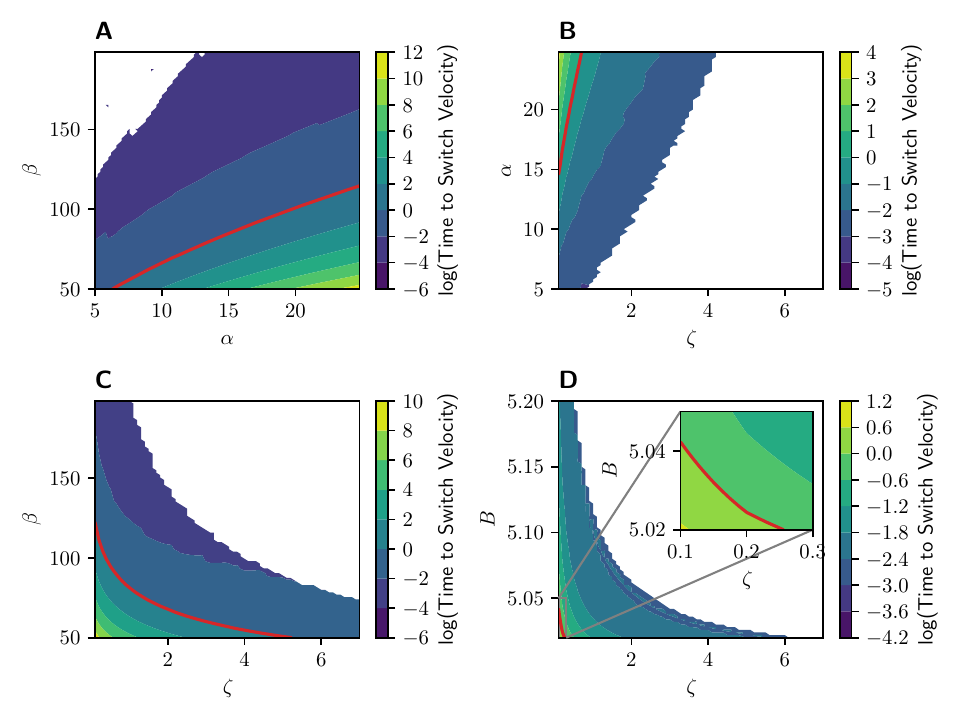}
    \caption{Average time to switch velocity as a function of various parameter pairs using the simplified master equation. The red contour indicates $\log(\text{Time to Switch Velocity}) = 0$. Remaining colors denote the natural logarithm of the mean time to switch velocity. A: $(\alpha,\beta)$, B: $(\zeta,\alpha)$, C: $(\zeta,\beta)$, D: $(\zeta,B)$. White regions regions denote parameter values for which the mean time to switch velocity cannot be computed, because there is only one peak in the steady-state distribution. Biologically-realistic parameter regimes are within an order of magnitude of the zero contour (red). Parameters that are not varied are kept at their default values: $n_D = n_U = 100$, $\zeta=\SI{1}{mg/s}$, $A=\SI{5}{nm}$, $B=\SI{5.05}{nm}$, $\alpha=\SI{14}{\s^{-1}}$, $\beta = \SI{126}{\s^{-1}}$.}
    \label{fig:v_pars}
\end{figure}

To better understand the effects of other molecular motor parameters on the mean time to switch vesicle velocity, we turn to Figure \ref{fig:v_pars}. Here, each panel shows $\tau$ as a contour plot as a function of parameter pairs. We select molecular motor parameters that may be regulated by the cell, e.g., by altering available levels of ATP \cite{maschi2021myosin} or through mechanical signaling mechanisms \cite{howard2009mechanical} that alter the spine geometry. Thus, we consider the parameter pairs $(\alpha,\beta)$, $(\zeta,\alpha)$, $(\zeta,\beta)$, and $(\zeta,B)$. The first pair, $(\alpha,\beta)$, is important to consider because attachment and detachment rates can greatly affect the probability to switch velocity. The remaining parameter pairs each involve $\zeta$, to better understand how the constriction geometry $\zeta$ and molecular motors affect the mean time to switch vesicle velocity.

Parameter regions corresponding to a single peak (for which the time to switch velocity cannot be computed) are shown in white. The red contour indicates where $\log(\text{Time to Switch Velocity})=0$. Biologically relevant parameter values are within one or two orders of magnitude of this curve. 

Note that the jaggedness of contours is due to the ambiguity present in peak detection. To compute $\tau$ for a given set of parameters, we compute the underlying steady-state distribution using \eqref{eq:master_steady} from Section \ref{sec:steady-state}, then look for peaks and their corresponding indices. We use a function called ``scipy.ndimage.maximum\_filter,'' a Python implementation that detects peaks in two-dimensional arrays. \yp{We chose to look for maxima that are locally maximal within a $3\times 3$ grid in the space of attached motors}. Once a collection of peaks is detected, we exclude peak sizes below \yp{a probability of} $\num{1e-5}$. In the vast majority of cases, we are left with 1, 2, or 3 peaks. Because the peak detection is performed on a $100\times100$ grid \yp{in the space of attached motor number}, where the peaks exist within a grid size of roughly $12\times12$, some ambiguity is inevitable, especially at the boundary where 2 peaks merge into 1, where there is a bifurcation. These sources of ambiguity are ultimately inconsequential -- the goal of capturing qualitative changes in $\tau$ as a function of molecular motor parameters is still achieved in Figure \ref{fig:v_pars}.

We start with straightforward observations about $\tau$ based on Figure \ref{fig:v_pars}. Panel A shows that the vesicle velocity will switch more rapidly for lesser $\alpha$ (greater $\beta$), and  will switch more slowly for greater $\alpha$ (lesser $\beta$). For increasing values of the constriction parameter $\zeta$ (Panels B--D), we consistently observe a decrease in the average time to switch velocities, consistent with prior results showing that increasing $\zeta$ pushes the bimodal steady-state peaks closer together until they merge to generate a stable zero-velocity solution (see Figure \ref{fig:v_switch} and prior work \cite{park2020dynamics,park2022coarse}). We will examine how these observations affect the probability and mean first passage time of vesicle translocation in the next section.

\section{Vesicle Translocation}\label{sec:mfpt}

\begin{figure}
    \centering
    \includegraphics[width=.5\textwidth]{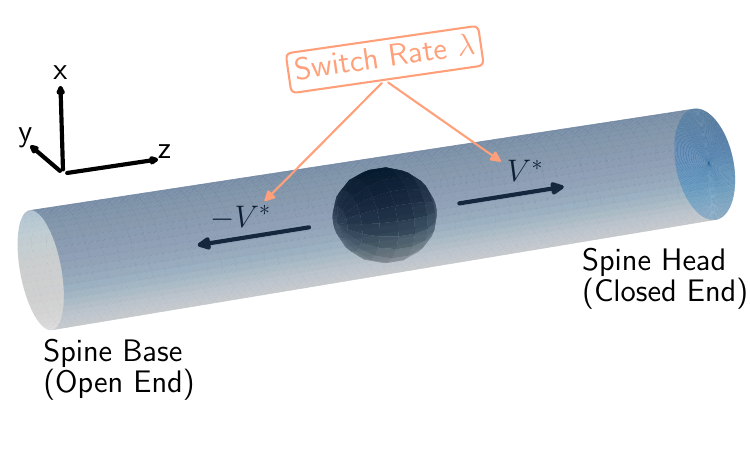}
    \caption{Cylindrical domain of the idealized dendritic spine and its relation to the telegraph process. The vesicle (black sphere) enters the spine base with an initial positive velocity $V^*$. The vesicle velocity switches between $\pm V^*$ at a rate given by $\lambda$.}\label{fig:domain}
\end{figure}

We idealize the problem of vesicle translocation as a rigid spherical object entering a cylindrical domain (Figure \ref{fig:domain}). This simplification captures the essential features of the original fluid dynamics problem: given that the diameter of the spherical object is only slightly smaller than the diameter of the cylinder, there is a large velocity gradient in the narrow space as the cytoplasmic fluid escapes towards the spine base. Here, lubrication theory applies, and the problem reduces to solving an ordinary differential equation. In particular, the degree of constriction is quantified by the confinement factor $\zeta$ \cite{fai2017active}. In general, the domain need not be cylindrical and the vesicle need not be spherical or rigid. However, these simplifying choices allow us to reframe the problem as the telegraph process.

\subsection{Telegraph Process}

The telegraph process is a continuous-time discrete-state stochastic process that switches between two distinct values with exponentially distributed waiting time \cite{kac1951some,goldstein1951diffusion}. The telegraph process has been analyzed and generalized in various directions under different names, e.g., Brownian motion processes \cite{fan2018random,rossetto2018one}, velocity-jump processes \cite{newby2011asymptotic}, correlated random walks \cite{weiss1984first,fan2018random,tang2019first}, run-and-tumble particles \cite{angelani2014first,angelani2015run,malakar2018steady,singh2020run}, and non-Markovian random walks \cite{hanggi1985first, masoliver1986first}. Thus, reframing our translocation problem in terms of the telegraph process allows us to utilize a rich history of mean first passage time results, as we did previously to arrive at the calculations in \cite{park2022coarse}, where we define the domain as the line segment $z\in[0,L]$ and calculate the mean first passage time out of only the $z=L$ boundary conditioned on the vesicle never reaching $z=0$. For brevity, we only state the equations to be used and refer the reader to \cite{park2022coarse} Section 4.1 for the full derivation.

We assume that the vesicle is initially located at $z=0$ with positive velocity $+V^*$. Then, the vesicle translocation probability $E$ and mean first passage time $S$ are given by,
\begin{align}
    E &= \frac{V^*}{ L/\tau(\zeta)+V^*},\label{eq:ep1}\\
    S &= \frac{1}{3} \left(\frac{L^2}{{\tau(\zeta)(V^*)}^2}+\frac{L}{L/\tau(\zeta)+V^*}+\frac{2 L}{V^*}\right),\label{eq:sp1}
\end{align}
where $\tau(\zeta)$ is the average waiting time to switch velocities, which can be rapidly computed using \eqref{eq:tauj}. 

\begin{figure}[ht!]
    \centering
    \includegraphics[width=\textwidth]{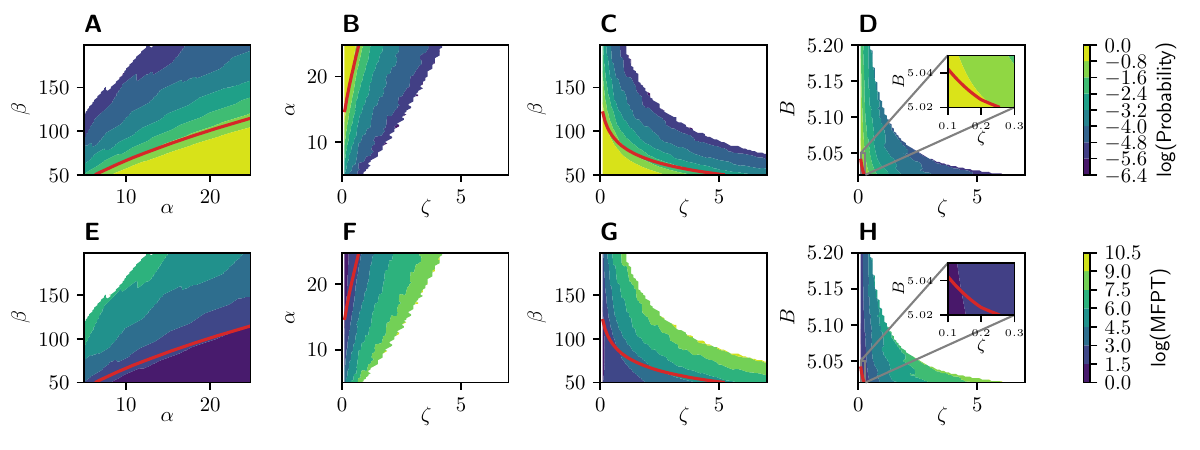}
    \caption{Probability (A--D) and mean first passage time (E--H) of translocation given a spine length of $L=\SI{200}{\nm}$. The starting position is always at the base of the spine. The red contour indicates $\log(\text{Time to Switch Velocity}) = 0$ from Figure \ref{fig:v_pars}. A,E: $(\alpha,\beta)$, B,F: $(\zeta,\alpha)$, C,G: $(\zeta,\beta)$, D,H: $(\zeta,B)$. White regions regions denote parameter values where the probability or mean first passage time cannot be computed due to the existence of only a single peak in the steady-state distribution. To avoid arbitrarily large mean first passage times, we exclude probabilities smaller than \num{1e-7} and mean first passage times (MFPTs) greater than \num{1e10} -- these thresholds do not meaningfully change the figure, because such extreme quantities only exist at a vanishingly small neighborhood near the boundary of the white regions. Parameters that are not varied are kept at their default values: $n_D = n_U = 100$, $\zeta=\SI{1}{mg/s}$, $A=\SI{5}{nm}$, $B=\SI{5.05}{nm}$, $\alpha=\SI{14}{\s^{-1}}$, $\beta = \SI{126}{\s^{-1}}$.}
    \label{fig:mfpt_200}
\end{figure}

Example outputs of \eqref{eq:ep1} and \eqref{eq:sp1} as a function of various molecular motor parameters are shown in Figure \ref{fig:mfpt_200} for a spine length of $L=\SI{200}{\nm}$. The top row (Panels A--D) shows $\log(E)$ for various parameter combinations, while the bottom row (Panels E--H) shows $\log(S)$ for various parameter combinations. We also generated the same figure for a spine length of $L=\SI{1000}{\nm}$ and found the results to be qualitatively similar -- see Appendix \ref{a:longer}. The jaggedness of the contours is inherited from Figure \ref{fig:v_pars}.

We begin with a few straightforward observations. Across all panels in Figure \ref{fig:mfpt_200}, the translocation probability is inversely proportional to the mean first passage time. This property is a consequence of the switching rates and our assumption that the vesicle starts at the spine base with an initial positive velocity towards the spine head. For low switching rates, the vesicle has a greater chance of unidirectional motion towards the vesicle head and therefore a greater likelihood of translocation. The mean first passage time tends to be lower as a consequence. At the opposite extreme, for high switching rates, the vesicle velocity resembles Brownian motion, and the probability of translocation becomes less likely because the probability of escaping through the base becomes more likely. For those vesicles that never escape through the base, their diffusive motion leads the vesicle to dwell for longer times within the cylinder before eventually reaching the spine head.



We now detail key observations and potential biological implications in each panel of Figure \ref{fig:mfpt_200}. In Panels A and B, we focus on the effects of altering attachment and detachment rates of myosin motors. These rates may be altered by ATP availability, where increased ATP availability tends to increase detachment and attachment rates simultaneously \cite{enrique2009kinetic}. Thus, we find that, for vesicles in spine necks of fixed width, ATP availability may have relatively little effect on vesicle translocation time and probability. The remaining panel pairs (B,F), (C,G), and (D,H) compare the effects of spine geometry to individual molecular motor parameters. We highlight these panels in the discussion because the effects of spine geometry on vesicle translocation are typically not discussed in the literature. Note that white regions denote zero velocity vesicle motion given our choice of equal numbers of up and down motors, but it is reasonable to assume that in reality there will exist a nonequal number of available up and down motors. In such cases, the vesicle will exhibit unidirectional motion determined by the dominant motor species and the velocity magnitude will be proportional to the difference in the number of available up and down motor species \cite{fai2017active,park2020dynamics}. 

In all remaining Figure \ref{fig:mfpt_200} Panels B--D and F--H, even moderate increases of the confinement factor $\zeta$ (corresponding to a thinner spine) are shown to significantly reduce the probability and mean first passage time to translocation, and thus reduce the spine's ability to maintain its postsynaptic density. This observation suggests an alternative mechanism for spine dysfunction, in addition to more established mechanisms such as a loss in spine density and an altered septin barrier in spine necks that are known to take place in patients with schizophrenia \cite{lee2015dendritic}. We note that in brain regions of schizophrenic patients in which spine densities are increased \cite{roberts2005synaptic}, dysfunction is still possible due to decreased vesicle trafficking from an excess of thin spines.

Figure \ref{fig:mfpt_200} Panels B and F suggest that increasing the molecular motor attachment rate $\alpha$ (by increasing ATP levels) could in principle alleviate the dysfunctional effect of thin spines. Figure \ref{fig:mfpt_200} Panels C and D at first glance suggest that lowering ATP levels to decrease motor detachment rates $\beta$ may potentially alleviate the dysfunctional effects of thin spines. However, this must be balanced with the corresponding decrease in $\alpha$ due to decreased ATP, and the corresponding decrease in vesicle translocation probability or increased translocation time. Thus, it is possible, if a neuron has an excess of thin spines, that there is no ATP level that can alleviate dysfunction. Finally, Figure \ref{fig:mfpt_200} Panels D and H show that translocation times and probabilities are sensitive to both the attachment position $B$ and confinement factor $\zeta$, with an especially strong sensitivity to $\zeta$. Note that $B-A$ is not the myosin motor step size, but the distance the myosin motor can stretch before detaching in its non-preferred direction (intuitively, how ``sticky'' the motor is when being stretched in its non-preferred direction). Thus, Panels D and H also show that forces from myosin motors that are attached to the vesicle but not undergoing a power stroke are likely negligible.




\section{Discussion}\label{sec:discussion}

In summary, we derived a simplified master equation from an agent-based model of vesicle transport. The simplified master equation yields analytically tractable quantities for the stationary distribution of attached motors (and therefore the stationary distribution of vesicle velocities) along with the vesicle velocity switching rate. These tractable quantities may be obtained from sparse linear systems and therefore take orders of magnitude less time to solve compared to our previous master equation in \cite{park2022coarse}, where we also solved a PDE of motor positions alongside the master equation. While using the PDE provided an exceptionally accurate approximation of the vesicle velocity while improving computational speed for large motor numbers, here by showing that the second moment of the density of motor positions is bounded and small relative to the mean motor positions (for biologically relevant parameter values), we are able to replace the position densities with their average values in the present study.

This simplification allows us to expand our prior work to much greater detail. In prior work, we fixed the mean waiting time to switch vesicle velocity $\tau$ and only varied the spine diameter and length, largely due to computational limitations \cite{park2022coarse}. Here, our simplification allows us to fix the cylinder length and width and instead explore how altering molecular motor parameters affected $\tau$ and, in turn, the time and probability of vesicle translocation. This opens the door to theoretical predictions about novel mechanisms of spine dysfunction, which we previously were unable to do.

From a biological perspective, one of the model predictions is that spine dysfunction may be caused, in part, by an excess of thin (but not necessarily long) spines, which can halt or greatly slow vesicle transport, in turn affecting synaptic function. This hypothesis is strikingly consistent with the literature. An excess of thin spines has been observed in people with intellectual disabilities \cite{purpura1974dendritic}. In Alzheimer's, the density of stubby spines is greatly reduced compared to control subjects \cite{boros2017dendritic} (however, it was observed that patients with Alzheimer's pathology maintain thin spines, while patients with Alzheimer's disorder do not, suggesting the importance of a balance of spine types). Excessive thin spines have been observed in fragile-X syndrome \cite{irwin2001abnormal}, with one 1985 study describing them as, ``long, thin, tortuous, dendritic spines'' \cite{rudelli1985adult}. Finally, mouse models of depression can show increases in the density of thin spines \cite{qiao2016dendritic}. However, further modeling studies and experiments are needed in order to validate the specific mechanisms suggested by this model.




\section{Acknowledgments}
The authors thank Sergei Pilyugin for helpful discussions about moment-generating functions.

\appendix

\section{Velocity Approximation Using Pad\'{e} Approximants}\label{a:velocity}

\begin{figure}[ht!]
	\centering
	\includegraphics[width=\textwidth]{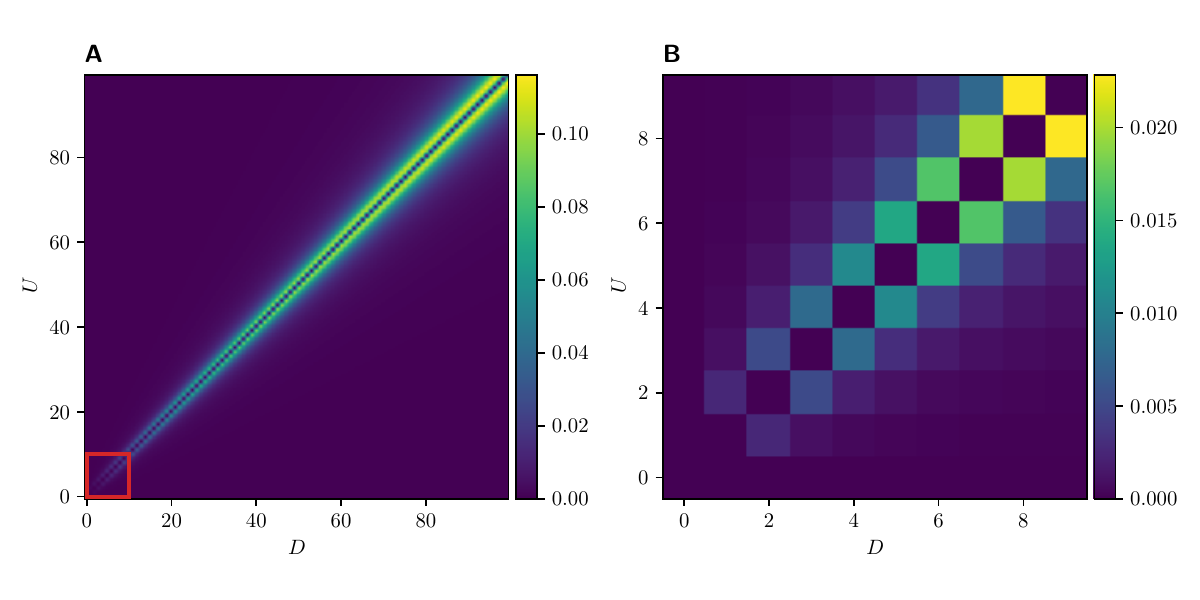}
	\caption{Plot of the relative velocity difference between the two velocity estimates \eqref{eqa:v1a} (the implicit velocity equation after using the mean attachment times) and \eqref{eqa:v2a} (the explicit velocity equation, solved after using Pad\'e approximants) for various numbers of attached motors. A: The relative error over all possible combinations of attached motors. B: The relative error when restricting the total $U$ and $D$ to the maximum mean attachment values (shown in Section \ref{sec:steady-state} following \eqref{eq:y_expression}). Parameters are $n_D=100$, $n_U=100$, $\alpha=14$, $\beta=126$, $A=5$, $B=5.1$, and $\zeta=0.2$.}\label{fig:dv}
\end{figure}

Recall that we approximated the true (transcendental) cargo velocity equation,
\begin{equation}\label{eqa:v1a}
	V_a = \frac{A(V_a-D)}{D(1-e^{\beta(A-B)/V_a})/\beta+ V_1/\beta + \zeta/k },
\end{equation}
in the case that $V_a > 0$, using a Pad\'{e} approximant, yielding:
\begin{equation}\label{eqa:v2a}
V_b = \frac{\beta ^2 \zeta  (A-B)+\sqrt{\mathcal{D}}+2 A \beta  k V_b-\beta  B k
		(D+V_b)}{2 (\beta  \zeta +k V_b)},
\end{equation}
where $V_b$ now satisfies a quadratic equation and can be solved explicitly. We claim that \eqref{eqa:v2a} serves as a reasonably good approximation by showing that the relative error between the two approximations, $|(V_a - V_b)/V_a|$, is small for a representative parameter set (Figure \ref{fig:dv}). In panel A, we find that the velocity differs by roughly 12\% when considering all possible combinations of attached motors for a representative parameter set. However, peaks in the steady-state distribution of this model only fluctuate within a relatively small range of motor attachments where the proportion of attached motors for both species is roughly bounded above by $\alpha/(\alpha+\beta)$ (see Section \ref{sec:steady-state} following \eqref{eq:y_expression}). The maximum relative error in this bounded region reaches 2.5\%. Thus, the velocity $V_b$ is a sufficiently close approximation to $V_a$.

\section{Mean First Passage Time for Longer Spines}\label{a:longer}

In Section \ref{sec:mfpt}, we calculated the probability and mean first passage time to vesicle translocation for a spine length of $L=\SI{200}{\nm}$. Here, we briefly describe the same quantities for a longer spine of length $L=\SI{1000}{\nm}$.

\begin{figure}[ht!]
    \centering
    \includegraphics[width=\textwidth]{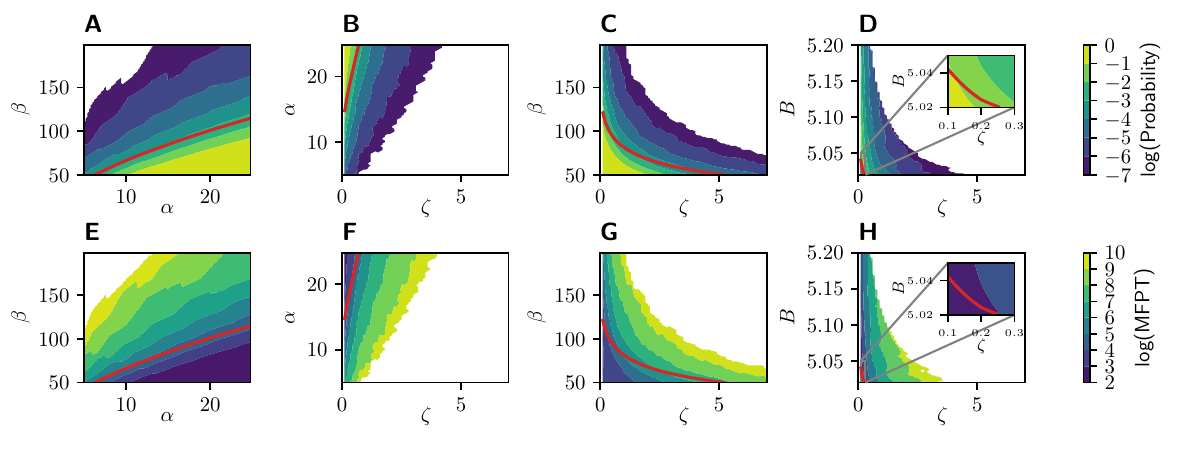}
    \caption{Probability (A--D) and mean first passage time (E--H) of translocation given a spine length of $L=\SI{1000}{\nm}$. The starting position is always at the base of the spine. The red contour indicates $\log(\text{Time to Switch Velocity}) = 0$. A,E: $(\alpha,\beta)$, B,F: $(\zeta,\alpha)$, C,G: $(\zeta,\beta)$, D,H: $(\zeta,B)$. White regions regions denote parameter values where the probability or mean first passage time cannot be computed because there exists only one peak in the steady-state distribution. To avoid arbitrarily large mean first passage times, we exclude probabilities smaller than \num{1e-7} and MFPTs greater than \num{1e10}.}
    \label{fig:mfpt_1000}
\end{figure}

\end{document}